\documentclass[journal]{IEEEtran}

\usepackage{amsmath,amsfonts,amssymb,mathtools}
\usepackage{algpseudocode}
\usepackage{algorithm}
\usepackage{array}
\usepackage[caption=false,font=normalsize,labelfont=sf,textfont=sf]{subfig}
\usepackage{textcomp}
\usepackage{stfloats}
\usepackage{url}
\usepackage{verbatim}
\usepackage{graphicx}
\usepackage{cite}
\usepackage{microtype}
\usepackage{placeins}
\usepackage{booktabs}
\usepackage{enumitem}
\usepackage{siunitx}
\usepackage{xcolor}
\usepackage{tabularx}
\usepackage{multirow}
\usepackage{tikz}
\usetikzlibrary{shapes.geometric, arrows.meta, positioning, calc, shadows.blur, backgrounds, fit}
\usepackage[colorlinks=true, allcolors=blue]{hyperref}
\usepackage{nomencl}
\makenomenclature

\newif\ifhighlighted
\highlightedfalse          % <-- change to \highlightedtrue for blue markup
\ifhighlighted
  \newcommand{\rev}[1]{\textcolor{blue}{#1}}
\else
  \newcommand{\rev}[1]{#1}
\fi

\DeclareSIUnit{\pu}{p.u.}
\DeclareSIUnit{\MW}{MW}

\definecolor{myblue}{RGB}{84, 153, 199}
\definecolor{myorange}{RGB}{243, 156, 18}
\definecolor{mygreen}{RGB}{82, 190, 128}
\definecolor{myred}{RGB}{231, 76, 60}
\definecolor{mygray}{RGB}{234, 236, 238}
\definecolor{mypurple}{RGB}{142, 68, 173}

\newcommand{\method}{AG2C\text{-}DT}
\newcommand{\fullmethod}{Autonomous Grid Generation Control with Decision Transformers}
\newcommand{\thead}[1]{\textbf{#1}}

\ifdefined\normalsize\else\newcommand{\normalsize}{\fontsize{10}{12}\selectfont}\fi

\begin{document}

\title{Generative Autonomous Grid Control: Integrating Decision Transformers with a Two-Stage Safety Stack}

\author{Mohamed Shamseldein,~\IEEEmembership{Senior Member,~IEEE}
	\thanks{M. Shamseldein is an Assistant Professor with the Department of Electrical Power and Machines, Faculty of Engineering, Ain Shams University, Cairo, Egypt.}}

\maketitle

% =====================================================================
%                           ABSTRACT
% =====================================================================
\begin{abstract}
\rev{The displacement of synchronous generation by inverter-based resources is accelerating power system frequency dynamics beyond the response capability of conventional automatic generation control. This paper presents Autonomous Grid Generation Control with Decision Transformers, a framework coupling an offline-trained Decision Transformer with a two-stage symbolic safety stack for secondary frequency control. The Decision Transformer learns a conditional dispatch policy from offline supervisory control and data acquisition records via sequence modeling, eliminating online exploration risks. A Constraint Verification Unit provides sub-ten-millisecond algebraic screening using real-time power transfer distribution factors, while an aggregate digital twin performs swing-equation-based dynamic stability certification. Validated on the Northeast Power Coordinating Council 140-bus system under low-inertia conditions, the proposed controller reduces the area control error integral by over 99\% relative to tuned automatic generation control, maintains a 59.4~Hz frequency nadir, and achieves inference latency of approximately 10~ms, well within real-time constraints. Comparative evaluation against a linear quadratic regulator baseline and structural analysis against conservative Q-learning demonstrate the advantages of the sequence-modeling formulation. Small-signal eigenvalue analysis characterizes the dominant 1.87~Hz electromechanical mode and confirms that the safety stack maintains stable operation across operating points. By falling back to tuned automatic generation control whenever proposals are rejected, the safety stack bounds worst-case performance to industry-standard levels in simulation.}
\end{abstract}


% =====================================================================
%                         NOMENCLATURE
% =====================================================================
\section*{Nomenclature}
\addcontentsline{toc}{section}{Nomenclature}
\begin{IEEEdescription}[\IEEEusemathlabelsep\IEEEsetlabelwidth{$\mathbf{P}_{line,t}$}]
	\item[$\mathcal{G}, \mathcal{L}, \mathcal{B}$] Sets of generators, transmission lines, and buses\rev{~[--]}.
	\item[$P_{g}, Q_{g}$] Active \rev{[MW]} and reactive \rev{[Mvar]} power generation of unit $g$.
	\item[$V_i, \theta_i$] Voltage magnitude \rev{[p.u.]} and phase angle \rev{[rad]} at bus $i$.
	\item[$\mathbf{V}_t, \mathbf{\theta}_t$] Vectors of all bus voltages \rev{[p.u.]} and angles \rev{[rad]} at time $t$.
	\item[$f_{agg,t}$] Aggregate system frequency at time $t$ \rev{[Hz]}.
	\item[$\mathbf{P}_{line,t}$] Vector of active power flows on all lines at time $t$ \rev{[MW]}.
	\item[$\mathbf{P}_{\text{gen},t}$] Generation dispatch vector at time $t$ \rev{[MW]}.
	\item[$\mathbf{P}_{\text{load},t}$] Load demand vector at time $t$ \rev{[MW]}.
	\item[$\Delta f$] System frequency deviation \rev{[Hz]}.
	\item[$ACE$] Area Control Error \rev{[MW]}.
	\item[$\mathcal{D}$] Offline dataset of trajectories \rev{[--]}.
	\item[$\pi_{\beta}$, $\pi_{\theta}$] Behavior (PID) policy and learned Decision Transformer policy \rev{[--]}.
	\item[$R_t$] Return-to-go at time step $t$ \rev{[--]}.
	\item[$K$] Context length for the Transformer window \rev{[timesteps]}.
	\item[$d_{model}$] Embedding dimension of the Transformer \rev{[--]}.
	\item[$\mathbf{H}$] Power Transfer Distribution Factor (PTDF) matrix \rev{[--]}.
	\item[$M, D$] Aggregated inertia \rev{[s]} and damping \rev{[p.u.]} constants.
	\item[$H_{sys}$] System-wide inertia constant \rev{[s]}.
	\item[$f_{min}$] Predicted frequency nadir \rev{[Hz]}.
	\item[$f_{UFLS}$] Under-Frequency Load-Shedding threshold \rev{[Hz]}.
	\item[$\Delta a_t$] Normalized action delta \rev{[--]}.
	\item[$\Delta P^{max}$] Maximum generator ramp per control step \rev{[MW]}.
	\item[$\theta_{OU}, \sigma_{OU}$] Ornstein--Uhlenbeck noise parameters \rev{[--]}.
	\rev{\item[$C_{op}$] Quadratic generation operating cost [\$/h].}
	\rev{\item[$\alpha_i, \beta_i, \gamma_i$] Generator cost curve coefficients [\$/MW$^2$h, \$/MWh, \$/h].}
	\rev{\item[$\lambda_f, \lambda_{ace}, \lambda_c, \lambda_v$] Reward weighting coefficients [--].}
	\rev{\item[$\zeta$] Damping ratio of electromechanical mode [--].}
	\rev{\item[$T_s$] Settling time [s].}
\end{IEEEdescription}

% =====================================================================
%                     SECTION I: INTRODUCTION
% =====================================================================
\section{Introduction}
\label{sec:intro}

\IEEEPARstart{T}{he} fundamental operational paradigm of power systems is shifting. The displacement of synchronous generation by Inverter-Based Resources (IBRs) has precipitated a decline in system inertia, accelerating frequency dynamics from the order of seconds to hundreds of milliseconds \cite{DOE2022,Milano2018}. Consequently, the timescale for corrective action is contracting beyond the cognitive limits of human operators, necessitating a transition toward higher degrees of automation. \rev{The U.S. Department of Energy has identified this ``inertia gap'' as a critical barrier to achieving renewable penetration above 60\% \cite{DOE2022}, while recent cascading events in low-inertia grids (e.g., the 2019 UK blackout) underscore the urgency of automated frequency response \cite{Kundur1994}.}

While Deep Reinforcement Learning (DRL) has demonstrated theoretical capability in handling fast dynamics \cite{Zhang2023}, its deployment in real-world control rooms remains virtually non-existent. This ``deployment gap'' stems from two fundamental flaws in standard online RL algorithms (e.g., Proximal Policy Optimization (PPO), Soft Actor-Critic (SAC)). First, they rely on stochastic exploration---trial-and-error learning---which poses unacceptable risks to critical infrastructure. Second, deep neural networks are opaque models lacking the rigorous, physics-based assurances required by Transmission System Operators (TSOs). \rev{Model-based approaches such as Linear Quadratic Regulators (LQR) and Model Predictive Control (MPC) offer analytical guarantees but assume linear time-invariant dynamics, which degrades under the nonlinear, time-varying conditions of modern low-inertia grids \cite{Sauer2017}.} \rev{A promising alternative is the Decision Transformer (DT) \cite{Chen2021}, which recasts control as supervised sequence modeling on existing operational logs. Crucially, DTs exhibit a ``stitching property'': by conditioning on a high target return, they synthesize optimal trajectories from sub-optimal demonstrations, enabling the agent to outperform the baseline controllers that generated the training data---without ever exploring online. For TSOs, this means leveraging years of archived SCADA data to train a policy that exceeds current AGC performance while remaining entirely offline and auditable.}

% --- MERGED RELATED WORK (formerly Section II) ---
\rev{\subsection{Related Work and Research Gaps}}

\rev{\subsubsection{Deep RL in Power Systems}}
The application of DRL to power system control has evolved from discrete action spaces (e.g., topological switching \cite{Zhang2023}) to continuous control of generator setpoints and inverter parameters. \rev{Huang \textit{et al.} \cite{Huang2020} applied deep deterministic policy gradient to voltage regulation on the IEEE 39-bus system, while Li \textit{et al.} \cite{Li2023} demonstrated SAC-based frequency control with promising laboratory results.} Complementary bibliometric analyses \cite{Miraftabzadeh2024} reinforce the need for controllers bridging academic prototypes and deployable solutions.

However, a significant limitation is the reliance on simplified test cases. The vast majority of DRL studies validate on small-scale benchmarks like the IEEE 39-bus or 118-bus systems \cite{IEEE_118}. Open-source environments such as Grid2Op \cite{Donnot2020} have accelerated development, but these models often lack inter-area oscillation modes inherent in regional backbones like the \textbf{NPCC 140-bus system}. This ``complexity gap'' raises doubts about transferability, mirroring data-access limitations highlighted by surveys of open grid datasets \cite{Aravena2025}.

\rev{\subsubsection{Offline Reinforcement Learning}}
Offline RL (or Batch RL) addresses the exploration risk by learning policies entirely from static, historical datasets. Standard approaches such as Conservative Q-Learning (CQL) \cite{Kumar2020} and Batch-Constrained Q-Learning (BCQ) \rev{\cite{Fujimoto2019}} rely on estimating value functions. However, these methods grapple with the ``deadly triad'' of instability: function approximation, bootstrapping, and off-policy distribution shift. When the agent queries Out-of-Distribution (OOD) actions, Q-value estimates can diverge, leading to dangerous policy artifacts \cite{Levine2020}. \rev{CQL mitigates this by adding a conservative penalty to Q-values for unseen actions, while BCQ constrains the policy to remain close to the behavior distribution. Both require careful hyperparameter tuning and can be overly conservative, under-utilizing the training data \cite{Levine2020}.}

\rev{\subsubsection{Sequence Modeling for Control}}
A recent paradigm treats RL as a sequence modeling problem \cite{Janner2021}. The Decision Transformer (DT) \cite{Chen2021} leverages the GPT architecture \cite{Vaswani2017} to predict actions autoregressively, conditioned on a target return-to-go. Unlike value-based methods, the DT avoids bootstrapping entirely, relying on supervised learning stability. By conditioning on a high target return, the DT extracts optimal sub-trajectories from sub-optimal datasets---the ``stitching property''---allowing the agent to outperform the demonstrators that generated the training data. \rev{Zheng \textit{et al.} \cite{Zheng2022} showed that online fine-tuning of DTs can further improve performance, while Yamagata \textit{et al.} \cite{Yamagata2023} extended DTs with Q-learning guidance for improved stitching.}

\rev{\subsubsection{Safe RL and Shielding}}
Ensuring safety in RL is typically approached via Constrained Markov Decision Processes (CMDPs), using Lagrangian relaxation to penalize violations during training. However, this offers no hard guarantees during inference. ``Shielding'' mechanisms \cite{Dalal2018} project unsafe actions onto a safe subspace, but most rely on simplified DC power flow or computationally expensive optimization violating real-time latency constraints. \rev{Recent work by Ozkan \textit{et al.} \cite{Ozkan2024} demonstrated RL-supported digital twin models for micro-grids, highlighting the promise of combining data-driven control with physics-based screening.}

\rev{\subsubsection{Literature Synthesis}}
\rev{Table~\ref{tab:lit_synthesis} synthesizes the reviewed approaches, highlighting their advantages, limitations, and the gaps addressed by the proposed framework.}

\begin{table*}[t]
\rev{
\centering
\caption{\rev{Synthesis of Related Approaches for Power System Frequency Control}}
\label{tab:lit_synthesis}
\resizebox{\textwidth}{!}{%
\begin{tabular}{@{}l l l l l@{}}
\toprule
\thead{Category} & \thead{Representative Methods} & \thead{Advantages} & \thead{Limitations} & \thead{Gap Addressed by AG2C-DT} \\
\midrule
Model-based control & LQR, MPC, H$_\infty$ & Analytical guarantees; well understood & Linear assumption; degrades under & Nonlinear sequence modeling \\
 & & & nonlinear/time-varying conditions & with physics-based safety stack \\
\midrule
Online RL & PPO \cite{Zhang2023}, SAC, & Handles nonlinear dynamics; & Exploration risk; no safety guarantees; & Offline training eliminates \\
 & DDPG \cite{Huang2020} & learns complex policies & validated only on small benchmarks & exploration risk entirely \\
\midrule
Offline RL (value-based) & CQL \cite{Kumar2020}, & No exploration needed; & Deadly triad; OOD divergence; & Supervised loss avoids \\
 & BCQ \cite{Fujimoto2019} & uses historical data & conservative bias & bootstrapping instability \\
\midrule
Sequence modeling & Decision Transformer & Stable supervised training; & No inherent safety mechanism; & Two-stage symbolic safety \\
(without safety) & \cite{Chen2021} & stitching property & limited to small-scale validation & stack + NPCC 140-bus validation \\
\midrule
Safe RL / Shielding & CMDP \cite{Dalal2018}, & Constraint awareness; & Soft penalties (no hard guarantees); & Deterministic ANDES-based \\
 & Barrier functions & formal frameworks available & simplified power flow models & symbolic Jacobian verification \\
\bottomrule
\end{tabular}%
}
}
\end{table*}

\rev{\subsection{Contributions}}

\rev{Autonomous Grid Generation Control with Decision Transformers (\method)} is introduced as a framework coupling a trajectory-sequence policy with a rigorous physics-based verification layer. The approach utilizes a DT \cite{Chen2021} to synthesize control policies from offline datasets. The environment leverages \textbf{ANDES} \cite{Cui2021}, a Python-based symbolic-numeric simulator, aligning with differentiable-simulator best practices \cite{Newbury2024}. ANDES enables encasing the generative policy in a high-fidelity \emph{Two-Stage Safety Stack} (TSSS) using symbolic Jacobians for precise verification.

\rev{The application domain is secondary frequency control (Automatic Generation Control) in bulk power systems, specifically targeting the redispatch of active power setpoints for dispatchable generators to maintain frequency and inter-area tie-line flow stability following load disturbances and contingencies. The framework operates at the Energy Management System (EMS) level, complementing---not replacing---primary governor response.}

The contributions are:
\begin{enumerate}
	\item \textbf{Offline Sequence Formulation:} Controller learning is cast as supervised sequence modeling on offline operational data, eliminating online exploration risks and Bellman backup instabilities.
	\item \textbf{Stitching Capability:} Empirical demonstration that the DT synthesizes optimal trajectories from sub-optimal, noisy PID demonstrations.
	\item \textbf{Realistic Validation (NPCC 140-bus):} Validation on the NPCC 140-bus system with complex inter-area oscillations, \rev{including systematic comparison against LQR and structural comparison against CQL/BCQ}.
	\item \textbf{Symbolic-Numeric Safety Stack:} Hierarchical verification using ANDES symbolic Jacobians ($<10$~ms) and aggregate swing-equation digital twin.
	\rev{\item \textbf{Stability Analysis:} Small-signal eigenvalue analysis characterizing the dominant inter-area mode, with inertia-level sensitivity evaluation demonstrating safety-stack robustness.}
\end{enumerate}

\rev{Table~\ref{tab:autonomy_levels} positions the proposed framework within a conceptual taxonomy of grid autonomy levels.}

\begin{table}[t]
	\centering
	\caption{Proposed Levels of Grid Autonomy}
	\label{tab:autonomy_levels}
	\resizebox{\columnwidth}{!}{%
		\begin{tabular}{@{}c l l@{}}
			\toprule
			\textbf{Level} & \textbf{Description} & \textbf{Power System Equivalent} \\
			\midrule
			\textbf{L0} & \textbf{Manual} & Phone dispatch; manual setpoints. \\
			\textbf{L1} & \textbf{Assisted} & AGC (reactive frequency/tie-line control). \\
			\textbf{L2} & \textbf{Partial} & RT-SCED / OPF (human approves setpoints). \\
			\textbf{L3} & \textbf{Conditional} & Black-box AI (human intervention for edge cases). \\
			\textbf{L4} & \textbf{High} & \textbf{\method.} Autonomous handling within \\
			& & design domain; safety stack screens violations; \\
			& & deterministic fallback if violation predicted. \\
			\textbf{L5} & \textbf{Full} & Fully autonomous under all conditions. \\
			\midrule
			\multicolumn{3}{@{}p{\dimexpr\columnwidth-2\tabcolsep}@{}}{\footnotesize\emph{Note:} Conceptual taxonomy; not a formal certification framework.} \\
			\bottomrule
		\end{tabular}%
	}
\end{table}

% =====================================================================
%              SECTION II: PROBLEM FORMULATION
% =====================================================================
\section{Problem Formulation}
\label{sec:formulation}

The autonomous grid control problem is formulated as a Markov Decision Process (MDP) defined by the tuple $(\mathcal{S}, \mathcal{A}, \mathcal{T}, \mathcal{R}, \gamma)$. Unlike static Optimal Power Flow (OPF), the objective is to learn a dynamic control policy $\pi(a_t|s_t)$ capable of stabilizing the nonlinear dynamics of the NPCC 140-bus system under stochastic disturbances.

\rev{\subsection{Dispatch Model}}
\rev{The underlying economic dispatch is formulated as a constrained optimization subproblem whose solution the DT learns to approximate from offline operational data. The full dispatch model is:}
\rev{
\begin{subequations}\label{eq:dispatch}
\begin{align}
\min_{P_{g_i}} \quad & \sum_{i \in \mathcal{G}} \left(\alpha_i P_{g_i}^2 + \beta_i P_{g_i} + \gamma_i\right) \label{eq:obj}\\
\text{s.t.} \quad & \sum_{i \in \mathcal{G}} P_{g_i} = \sum_{j \in \mathcal{B}} P_{d_j} + P_{\text{loss}} \label{eq:balance}\\
& P_{g_i}^{\min} \leq P_{g_i} \leq P_{g_i}^{\max}, \quad \forall i \in \mathcal{G} \label{eq:gen_lim}\\
& |P_{l}| \leq P_{l}^{\max}, \quad \forall l \in \mathcal{L} \label{eq:line_lim}\\
& V_b^{\min} \leq V_b \leq V_b^{\max}, \quad \forall b \in \mathcal{B} \label{eq:volt_lim}\\
& |P_{g_i}(t) - P_{g_i}(t-1)| \leq \Delta P_{g_i}^{\max}, \quad \forall i \in \mathcal{G} \label{eq:ramp}\\
& |ACE| \leq ACE^{\max} \label{eq:ace_lim}
\end{align}
\end{subequations}
where \eqref{eq:obj} minimizes quadratic generation cost with coefficients $(\alpha_i, \beta_i, \gamma_i)$ extracted from ANDES cost curves; \eqref{eq:balance} enforces active power balance including losses; \eqref{eq:gen_lim}, \eqref{eq:line_lim}, \eqref{eq:volt_lim}, and \eqref{eq:ramp} enforce generator capacity, thermal line, voltage, and ramping limits; and \eqref{eq:ace_lim} bounds the Area Control Error per NERC BAL-003 \cite{NERC_BAL}. The DT does not solve \eqref{eq:dispatch} explicitly but learns dispatch trajectories from the behavior policy that approximately satisfy these constraints; the safety stack then certifies hard constraint satisfaction at deployment.}

\subsection{State and Action Spaces}
The state space $\mathcal{S} \subset \mathbb{R}^{N_s}$ captures both local nodal conditions and global inter-area states. Using ANDES, the state vector $s_t$ of dimension $N_s \approx 700$ comprises:
\begin{equation}
	s_t = [ \mathbf{V}_t, \mathbf{\theta}_t, f_{agg,t}, \mathbf{P}_{line,t}, \mathbf{P}_{gen,t}, \mathbf{P}_{load,t}, ACE_t ]^\top
\end{equation}
where $\mathbf{V}_t, \mathbf{\theta}_t \in \mathbb{R}^{140}$ represent voltage magnitudes and phase angles at every bus, and $\mathbf{P}_{line,t} \in \mathbb{R}^{233}$ captures real-time power flows including critical tie-lines linking the NYISO, ISO-NE, and IESO equivalents.

The action space $\mathcal{A} \subset \mathbb{R}^{N_a}$ consists of active power setpoints for the dispatchable fleet ($N_a = 48$):
\begin{equation}
	a_t = [ P_{g_1}^{set}, P_{g_2}^{set}, \dots, P_{g_{48}}^{set} ]^\top
\end{equation}
Actions are continuous and bounded by ramping rates ($\Delta P_g^{max}$) and capacity limits, enforced via the Safety Stack.

\subsection{Reward Structure}
The reward function $r(s_t, a_t)$ in \eqref{eq:reward} aligns agent behavior with NERC BAL-003 frequency response standards \cite{NERC_BAL}:
\begin{equation} \label{eq:reward}
	r_t = - \left( \lambda_{f} |\Delta f_t| + \lambda_{ace} |ACE_t| + \lambda_{c} C_{op}(a_t) + \lambda_{v} \Psi(s_t) \right)
\end{equation}
with $\lambda_f = \lambda_{ace} = 1.0$, $\lambda_c = 0.1$, and $\lambda_v = 10.0$, so that frequency control objectives dominate economic dispatch. \rev{The economic dispatch term $C_{op}(a_t) = \sum_i \alpha_i a_{t,i}^2 + \beta_i a_{t,i} + \gamma_i$ uses ANDES cost curve coefficients.} The hinge penalty
\begin{equation}
	\Psi(s_t) = \sum_{l \in \mathcal{L}} \max(0, |P_{l,t}| - P_{l}^{\max}) + \sum_{b \in \mathcal{B}} \max(0, |V_{b,t} - 1.0| - \delta_v)
\end{equation}
regularizes offline training; at deployment the safety stack provides hard enforcement.

\subsection{Offline RL Objective}
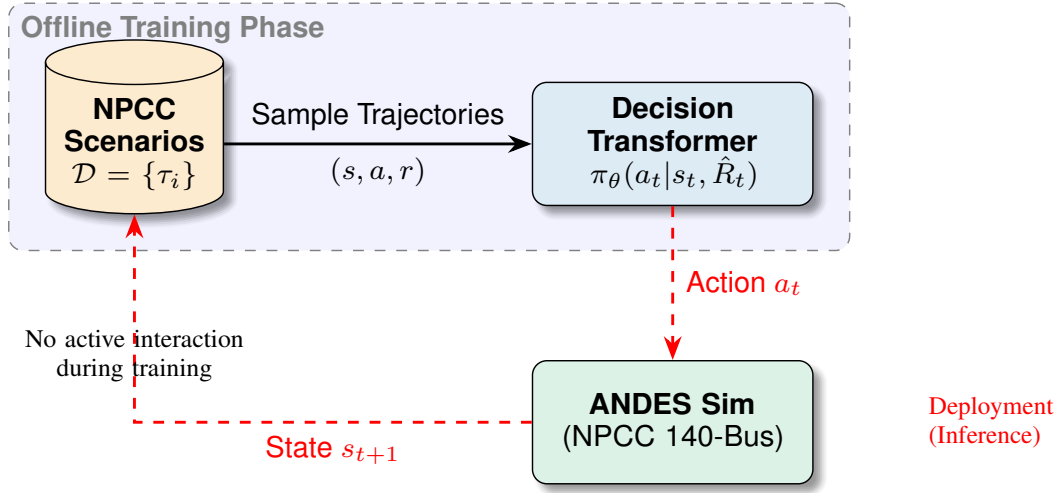
\begin{figure*}[t]
	\centering
	\resizebox{0.8\textwidth}{!}{%
		\begin{tikzpicture}[
			node distance=1.5cm, auto, font=\sffamily\footnotesize,
			block/.style = {rectangle, draw, fill=white, text width=2.5cm, text centered, rounded corners, minimum height=1.2cm, blur shadow},
			cloud/.style = {ellipse, draw, fill=gray!10, text width=2.0cm, text centered, minimum height=1.0cm, blur shadow},
			database/.style = {cylinder, draw, shape border rotate=90, aspect=0.25, fill=white, text width=1.5cm, text centered, minimum height=1.5cm, blur shadow},
			line/.style = {draw, -{Stealth}, thick},
			dashed_box/.style = {draw=gray, dashed, rounded corners, fill=blue!5, inner sep=10pt}
			]
			\node [database, fill=myorange!20] (dataset) {\textbf{NPCC Scenarios}\\ $\mathcal{D} = \{\tau_i\}$};
			\node [block, right=of dataset, xshift=1.5cm, fill=myblue!20] (policy) {\textbf{Decision Transformer}\\ $\pi_\theta(a_t | s_t, \hat{R}_t)$};
			\node [block, below=of policy, xshift=0cm, fill=mygreen!20] (environment) {\textbf{ANDES Sim}\\ (NPCC 140-Bus)};
			\begin{scope}[on background layer]
				\node [dashed_box, fit=(dataset) (policy), label={[anchor=north west, text=gray]north west:\textbf{Offline Training Phase}}] (offline_box) {};
			\end{scope}
			\path [line] (dataset) -- node [above] {Sample Trajectories} node [below] {$(s, a, r)$} (policy);
			\path [line, dashed, red] (policy) -- node [right] {Action $a_t$} (environment);
			\path [line, dashed, red] (environment) -| node [near start, below] {State $s_{t+1}$} (dataset);
			\node [below=of dataset, yshift=0.5cm, text width=3cm, align=center, font=\scriptsize] {No active interaction\\during training};
			\node [right=of environment, xshift=-0.5cm, font=\scriptsize, text=red, align=left] {Deployment\\(Inference)};
		\end{tikzpicture}%
	}
	\caption{The Offline Learning Paradigm. The agent ($\pi_\theta$) optimizes the policy using a static dataset ($\mathcal{D}$) generated from the ANDES simulator. Physical grid interaction occurs only during deployment.}
	\label{fig:offline_rl_diagram}
\end{figure*}

A static dataset $\mathcal{D} = \{ \tau_i \}_{i=1}^N$ is collected from a behavior policy $\pi_\beta$ (baseline PID controller within ANDES). The objective is to learn $\pi_\theta$ maximizing $J(\pi_\theta) = \mathbb{E}_{\tau \sim \pi_\theta}[\sum \gamma^t r_t]$ using only $\mathcal{D}$, avoiding online exploration risks \cite{Chen2025}. \rev{Figure~\ref{fig:offline_rl_diagram} summarizes this separation between offline training and deployment-time grid interaction.}

% =====================================================================
%              SECTION III: THE AG2C-DT FRAMEWORK
% =====================================================================
\section{The \method~Framework}
\label{sec:framework}

The framework consists of two integrated components: (1) a Generative Control Agent based on the Decision Transformer architecture, and (2) a deterministic TSSS implemented within the ANDES environment.

\subsection{Generative Control via Decision Transformers}
Grid control is modeled as a conditional sequence generation problem. A trajectory is represented as:
\begin{equation}
	\tau = (\hat{R}_1, s_1, a_1, \hat{R}_2, s_2, a_2, \dots, \hat{R}_T, s_T, a_T)
\end{equation}
where $\hat{R}_t = \sum_{t'=t}^T r_{t'}$ is the Return-to-Go (RTG). The DT learns $\pi_\theta$ parameterized by a GPT-style causal transformer.

\subsubsection{Input Representation}
At timestep $t$, the model receives a context window of the last $K$ steps ($K=32$, i.e., 3.2~s of history) to capture both governor and inter-area dynamics. Input tokens are projected into an embedding space of dimension $d_{model}=256$ with a learned timestep embedding.

\subsubsection{Autoregressive Prediction}
The model predicts the next action via causal self-attention:
\begin{equation}
	a_t \sim P_\theta(a_t | \hat{R}_{t-K:t}, s_{t-K:t}, a_{t-K:t-1})
\end{equation}
The DT outputs \emph{normalized action deltas} $\Delta a_t = (a_t - a_{t-1}) / \sigma_a$ rather than absolute setpoints. The supervised loss combines an MSE term on $\Delta a_t$ with a ramp-regularization penalty $\lambda_{\text{ramp}}\|\Delta a_t\|_2^2$. At inference, the predicted delta is denormalized, smoothed with an exponential moving-average filter, clipped to the configured ramp budget (300~MW), and integrated with the last ANDES dispatch. This delta parameterization removed gigawatt-scale jumps observed in early prototypes and allowed the safety stack to approve virtually all DT proposals.

\begin{figure}[t]
	\centering
	\includegraphics[width=\columnwidth]{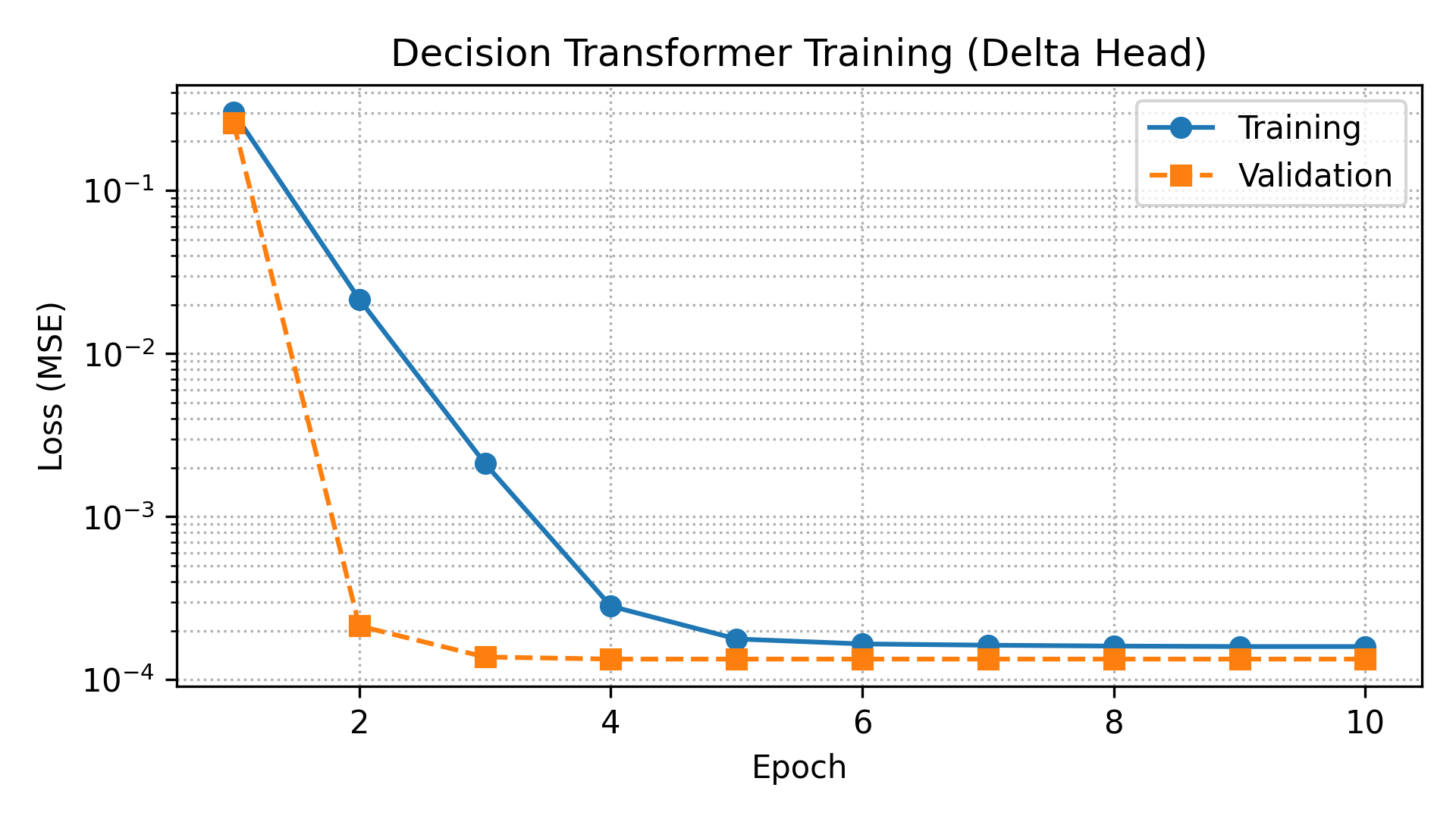}
	\caption{Training dynamics of the delta-based Decision Transformer. Validation loss plateaus after four epochs, motivating early stopping.}
	\label{fig:training_curve}
\end{figure}

During inference, the model is prompted with $\hat{R}_{target} > \hat{R}_{expert}$, conditioning the agent to generate trajectories yielding higher performance than the training data average (the ``stitching'' property). \rev{The resulting optimization trajectory is shown in Fig.~\ref{fig:training_curve}.}

\subsection{The Two-Stage Safety Stack via ANDES}
\label{sec:safety_stack}
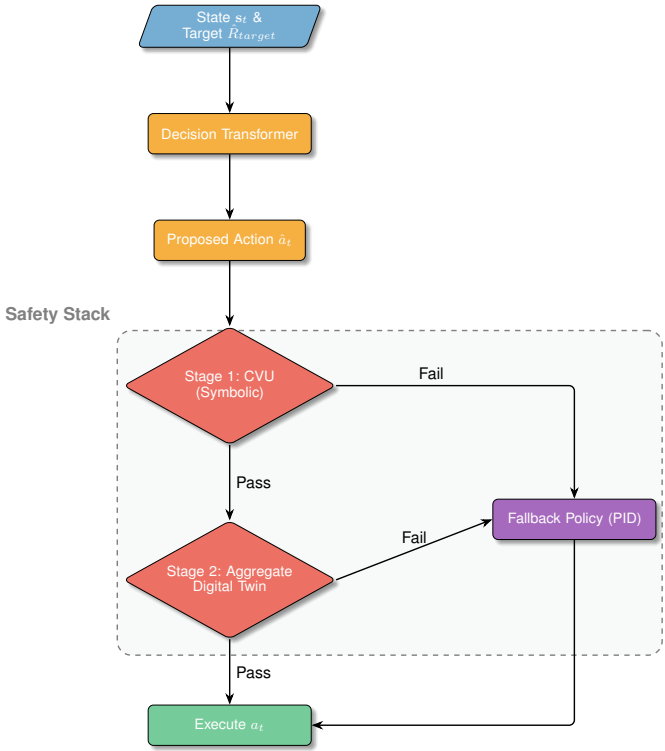
\begin{figure}[htbp]
	\centering
	\resizebox{\columnwidth}{!}{%
		\begin{tikzpicture}[
			io/.style = {trapezium, trapezium left angle=70, trapezium right angle=110, minimum width=3cm, minimum height=0.8cm, text centered, draw=black, fill=myblue!80, text=white, rounded corners=3pt, blur shadow={shadow xshift=0.5ex, shadow yshift=-0.5ex}, font=\sffamily\footnotesize, text width=2.8cm},
			process/.style = {rectangle, minimum width=3cm, minimum height=0.8cm, text centered, text width=3cm, draw=black, fill=myorange!80, text=white, rounded corners=3pt, blur shadow={shadow xshift=0.5ex, shadow yshift=-0.5ex}, font=\sffamily\footnotesize},
			decision/.style = {diamond, aspect=1.8, minimum width=2.5cm, minimum height=0.8cm, text centered, draw=black, fill=myred!80, text=white, rounded corners=3pt, blur shadow={shadow xshift=0.5ex, shadow yshift=-0.5ex}, font=\sffamily\footnotesize, text width=2.5cm},
			output_node/.style = {rectangle, minimum width=3cm, minimum height=0.8cm, text centered, text width=3cm, draw=black, fill=mygreen!80, text=white, rounded corners=3pt, blur shadow={shadow xshift=0.5ex, shadow yshift=-0.5ex}, font=\sffamily\footnotesize},
			fallback_node/.style = {rectangle, minimum width=3cm, minimum height=0.8cm, text centered, text width=3cm, draw=black, fill=mypurple!80, text=white, rounded corners=3pt, blur shadow={shadow xshift=0.5ex, shadow yshift=-0.5ex}, font=\sffamily\footnotesize},
			arrow/.style = {thick, -{Stealth}, rounded corners=3pt},
			stage_box/.style = {draw=gray, thick, dashed, rounded corners=8pt, fill=mygray, fill opacity=0.3},
			node distance=1.3cm, align=center, font=\sffamily\small
			]
			\node (input) [io] {State $\mathbf{s}_t$ \& Target $\hat{R}_{target}$};
			\node (model) [process, below=of input] {Decision Transformer};
			\node (action) [output_node, below=of model, fill=myorange!80, text width=2.5cm] {Proposed Action $\hat{a}_t$};
			\node (cvu) [decision, below=of action] {Stage 1: CVU (Symbolic)};
			\node (sim) [decision, below=of cvu, yshift=-0.2cm] {Stage 2: Aggregate Digital Twin};
			\node (fallback) [fallback_node, right=of sim, xshift=1.8cm, yshift=1.2cm] {Fallback Policy (PID)};
			\node (output) [output_node, below=of sim] {Execute $a_t$};
			\draw [arrow] (input) -- (model);
			\draw [arrow] (model) -- (action);
			\draw [arrow] (action) -- (cvu);
			\draw [arrow] (cvu) -- node[right] {Pass} (sim);
			\draw [arrow] (sim) -- node[right] {Pass} (output);
			\draw [arrow] (cvu.east) -| node[above, pos=0.2] {Fail} (fallback.north);
			\draw [arrow] (sim.east) -- node[above] {Fail} (fallback.west);
			\draw [arrow] (fallback.south) |- (output);
			\begin{scope}[on background layer]
				\node [stage_box, fit=(cvu) (sim) (fallback), yshift=-0.2cm] (sandbox_box) {};
			\end{scope}
			\node[above left, font=\sffamily\bfseries, text=gray] at (sandbox_box.north west) {Safety Stack};
		\end{tikzpicture}
	}
	\caption{Two-Stage Safety Stack Architecture. Stage~1 uses ANDES symbolic Jacobians for linearized sensitivity checks. Stage~2 uses an aggregate swing-equation twin for dynamic screening.}
	\label{fig:sandbox_flowchart}
\end{figure}

Deploying a generative model in critical infrastructure requires a fail-safe mechanism. A hierarchical safety stack leveraging the ANDES hybrid symbolic-numeric framework acts as a deterministic filter (Fig.~\ref{fig:sandbox_flowchart}).

\rev{\subsubsection{The ANDES Simulation Environment}}
\rev{ANDES \cite{Cui2021} is a Python-based open-source power system simulation tool that combines symbolic equation derivation with high-performance numerical solvers. It was selected for this work for three reasons: (1)~it provides symbolic Jacobian matrices that can be extracted and evaluated at any operating point without finite-difference approximation, enabling the sub-millisecond sensitivity computations required by the CVU; (2)~it ships with validated dynamic models including the NPCC 140-bus benchmark with full generator dynamic data (inertia constants, damping coefficients, exciter and governor models); and (3)~its Python-native interface permits seamless integration with PyTorch-based ML pipelines. The symbolic engine constructs the system differential-algebraic equations (DAEs) from component-level models and derives $\partial \mathbf{f}/\partial \mathbf{x}$ and $\partial \mathbf{g}/\partial \mathbf{y}$ analytically, which are then evaluated numerically at the current operating point.}

\subsubsection{Stage 1: Constraint Verification Unit (CVU)}
The CVU provides ultra-fast ($<10$~ms, measured $\approx 2$~ms) verification of algebraic constraints. ANDES extracts the instantaneous system Jacobian $\mathbf{J}_{sys}$ symbolically, enabling dynamic computation of the PTDF matrix $\mathbf{H}$:
\begin{equation}
	\mathbf{P}_{line}^{new} \approx \mathbf{P}_{line}^{meas} + \mathbf{H}(\mathbf{J}_{sys}) \cdot \Delta \mathbf{P}_{gen}
\end{equation}
A participation matrix $\mathbf{A}\in\mathbb{R}^{n_{\text{bus}}\times n_{\text{gen}}}$ maps generator deltas to bus injections with $\mathbf{A}^{\top}\mathbf{1}=\mathbf{1}$ preserving net power. Stage~1 clips generator ramps to $\pm300$~MW per 0.1~s, checks 0.95--1.05~p.u.\ voltages, and compares predicted flows against NPCC emergency ratings (600--1400~MW).

\subsubsection{Stage 2: Fast Digital Twin}
Actions passing the CVU enter dynamic screening. Full DAE simulation of the 48-generator system is computationally prohibitive for real-time control. Instead, an aggregate swing reduction is employed. ANDES aggregates inertia and damping:
\begin{equation}
	\frac{2H_{sys}}{\omega_0} \frac{d \Delta f}{dt} = \Delta P_{m} - \Delta P_{e} - D_{sys} \Delta f
\end{equation}
\rev{where $H_{sys} = \sum_i H_i S_i / \sum_i S_i$ is the capacity-weighted system inertia [s] and $D_{sys}$ is the aggregate damping [p.u.].} This reduced-order model predicts $f_{min}$ over a 1.0~s horizon \rev{in $<$1~ms}. If the predicted nadir triggers UFLS ($f_{UFLS} = 59.4$~Hz) or exceeds the 0.5~Hz/s Rate of Change of Frequency (ROCOF) cap, the action is rejected. Because the aggregate reduction collapses the system into a single equivalent machine, Stage~2 does not detect inter-area oscillatory modes; it strictly protects global frequency/ROCOF limits. \rev{This limitation is quantified through eigenvalue analysis in Section~\ref{sec:eigenvalue}.}

\rev{Combined with transformer inference ($\approx$10~ms), the entire propose-verify cycle averages $\approx$10~ms per step, well within the 4~s BAL-003 telemetry budget.}

\subsection{Inference Algorithm and Fallback}
If either the CVU or the Digital Twin rejects the DT's action, control reverts to baseline PID (AGC) logic, bounding worst-case performance to current industry standards (Algorithm~\ref{alg:inference}).

\begin{algorithm}
	\caption{\method~Inference Loop with Safety Stack}
	\label{alg:inference}
	\begin{algorithmic}[1]
		\State \textbf{Input:} State $s_t$, History buffer $\mathcal{B}_{hist}$, Target $\hat{R}_{target}$
		\State Update $\mathcal{B}_{hist} \leftarrow \mathcal{B}_{hist} \cup \{s_t\}$
		\State $\hat{a}_t \leftarrow \text{DT}_\theta(\text{Context}(\mathcal{B}_{hist}), \hat{R}_{target})$
		\State $\mathbf{P}_{line}^{pred} \leftarrow \mathbf{P}_{line}^{meas} + \mathbf{H} (\hat{a}_t - a_{t-1})$
		\If{$\max(|\mathbf{P}_{line}^{pred}|) > P^{\max}$}
			\State \textbf{return} $a_{PID}(s_t)$ \Comment{CVU rejection}
		\EndIf
		\State $f_{traj} \leftarrow \text{SimulateSwing}(\hat{a}_t, s_t)$
		\If{$\min(f_{traj}) < f_{UFLS}$}
			\State \textbf{return} $a_{PID}(s_t)$ \Comment{Digital Twin rejection}
		\EndIf
		\State \textbf{return} $\hat{a}_t$
	\end{algorithmic}
\end{algorithm}

% =====================================================================
%           SECTION IV: EXPERIMENTAL SETUP
% =====================================================================
\section{Experimental Setup}\label{sec:experiments}

\subsection{NPCC 140-Bus System}
The framework is validated on the NPCC 140-bus system, the publicly distributed benchmark shipped with ANDES. This reduced-order model represents the interconnected backbone of the Northeast Power Coordinating Council (NPCC), comprising 48 synchronous generators and 233 transmission lines across five areas corresponding to NYISO, ISO-NE, IESO, PJM, and MISO equivalents.

\rev{\subsection{Low-Inertia Modification and Generator Selection Criteria}}
\rev{To test robustness under modern grid conditions, 14 synchronous generators were replaced with Type-4 wind turbine models (\texttt{REGCA1}) from the ANDES library. The selection followed a deterministic capacity-based criterion: generators were ranked by rated capacity ($S_n$) in descending order, and the top 30\% by count (14 out of 48) were selected for replacement. Because the largest-rated units are chosen, this fraction represents a disproportionately large share of installed capacity, reflecting realistic utility practice where the largest conventional units are most likely to be displaced by renewable generation due to economic merit-order effects. Specifically, the inertia constants of the selected generators were reduced to 10\% of their original values (with a floor of $H = 0.05$~s), simulating the near-zero inertia contribution of grid-following inverters.}

\rev{This modification reduces $H_{sys}$ from approximately 5.2~s to 3.1~s and exacerbates the lightly damped inter-area oscillation mode between NYISO and ISO-NE (approximately 1.8--2.2~Hz). To assess sensitivity, results are reported for multiple inertia levels in Section~\ref{sec:ablation}.}

\rev{\subsection{Impact of Varying Assumptions}}
\rev{To address sensitivity to the low-inertia assumption, four inertia reduction levels are evaluated: 0\% (original NPCC), 15\%, 30\% (baseline), and 50\%. The results of these variations are presented in Table~\ref{tab:ablation_inertia}. Operating-point sensitivity is evaluated by sweeping the system load from 0.8$\times$ to 1.2$\times$ of the base case; both controllers maintain frequency at 60~Hz across all load levels, while the DT intervention rate remains near 1\% at nominal and high-load points (1.0$\times$--1.2$\times$) but rises sharply at lighter-load conditions (0.8$\times$--0.9$\times$), where PID fallback safely handles the mismatch.}

\subsection{Training Data Generation}
\rev{The offline dataset was generated by simulating the baseline PID controller over 12~hours of operation (432{,}000 timesteps at $\Delta t = 0.1$~s), stored in HDF5 format. The data source is entirely synthetic, generated via the ANDES/PYPOWER simulation environment using the NPCC 140-bus case with the low-inertia modification described above.} Ornstein--Uhlenbeck noise ($\theta_{OU}=0.15$, $\sigma_{OU}=0.2$) was injected into PID actions to ensure diverse, sub-optimal behaviors necessary for the stitching property. \rev{Domain randomization was applied across chunks: load scaling $\in [0.85, 1.15]$, system inertia $\in [3, 8]$~s, damping $\in [0.5, 1.5]$~p.u., and PID gains swept over $k_p \in [0.3, 0.9]$, $k_i \in [0.03, 0.2]$, $k_d \in [0.01, 0.08]$.}

The dataset was split into 345{,}575 training, 43{,}196 validation, and 43{,}198 test sequences of length $K=32$. \rev{A dataset-size ablation (Section~\ref{sec:ablation_dataset}) confirms that as little as 10\% of the training data yields a functional policy, with the TSSS ensuring safe operation at all data fractions.}

\paragraph*{Baseline PID Settings}
The fallback and evaluation PID controllers use $(k_p, k_i, k_d) = (0.5, 0.1, 0.05)$ with ACE bias $= 10$~MW/0.1~Hz per NERC BAL-003.

\rev{\subsection{Baseline Controllers}}
\rev{\subsubsection{PID/AGC Baseline}
The tuned PID controller serves as the primary baseline, representing current industry-standard AGC. It operates deterministically during evaluation without injected noise.}

\rev{\subsubsection{LQR Baseline}
A Linear Quadratic Regulator is constructed by linearizing the power system dynamics around the nominal operating point. The state matrix $\mathbf{A}_{sys}$ is derived from the swing equation and PTDF-based power flow sensitivities, while the input matrix $\mathbf{B}$ maps generator setpoint changes to state variations. The LQR gain $\mathbf{K}^*$ is obtained by solving the continuous-time algebraic Riccati equation:
\begin{equation}
	\mathbf{A}^\top \mathbf{P} + \mathbf{P}\mathbf{A} - \mathbf{P}\mathbf{B}\mathbf{R}^{-1}\mathbf{B}^\top \mathbf{P} + \mathbf{Q} = 0
\end{equation}
with $\mathbf{Q}$ penalizing frequency deviation and ACE, and $\mathbf{R}$ penalizing control effort. This provides a principled model-based comparison that highlights where the DT's nonlinear sequence modeling offers advantages over linear control theory.}

\rev{\subsubsection{Structural Comparison with CQL and BCQ}
Direct empirical comparison with CQL \cite{Kumar2020} and BCQ \cite{Fujimoto2019} on the NPCC 140-bus system remains computationally intensive and is deferred to future work. However, a structural comparison is provided in Table~\ref{tab:cql_comparison}, analyzing the algorithmic properties, failure modes, and suitability of each approach for the safety-critical grid control domain. The key advantage of the DT formulation is the elimination of bootstrapping: CQL and BCQ both require iterative Bellman updates that can diverge when the 703-dimensional state space induces significant OOD queries, whereas the DT uses a stable supervised MSE loss on action deltas.}

\begin{table}[t]
\rev{
\centering
\caption{\rev{Structural Comparison of Offline RL Approaches}}
\label{tab:cql_comparison}
\resizebox{\columnwidth}{!}{%
\begin{tabular}{@{}l c c c@{}}
\toprule
\thead{Property} & \thead{CQL} & \thead{BCQ} & \thead{\method} \\
\midrule
Training objective & Q-learning & Constrained Q & Supervised MSE \\
Bootstrapping & Yes & Yes & No \\
OOD divergence risk & Moderate & Low & None \\
Conservative bias & High & Moderate & None \\
Stitching capability & Limited & Limited & Strong \\
Safety integration & Requires wrapper & Requires wrapper & Native (TSSS) \\
Training stability & Sensitive to $\alpha$ & Sensitive to threshold & Stable \\
\bottomrule
\end{tabular}%
}
}
\end{table}

\paragraph*{Implementation Details}
\rev{Random seeds are fixed for data generation (42) and evaluation disturbances (31415); model weight initialization is not seeded.} The DT uses 6 layers, 8-head self-attention, $d_{model}=256$, GELU activations, and 0.1 dropout. Training runs for \rev{5} epochs with batch size 64 using AdamW ($lr = 3\times10^{-4}$, weight decay $= 0.01$) on an AMD Ryzen 7 CPU with NVIDIA RTX 4070 GPU ($\approx$20 min/epoch). \rev{Table~\ref{tab:computational} compares computational requirements with related approaches.}

% =====================================================================
%              SECTION V: RESULTS AND ANALYSIS
% =====================================================================
\section{Results and Analysis}\label{sec:results}

\rev{Agents are evaluated over 22 episodes of 400 steps each (40~s duration) under random load step disturbances with deterministic seed 31415 for reproducibility; 20 base scenarios are drawn from the randomised disturbance schedule and 2 stress-test scenarios (tie-line trip and extreme load step) exercise the safety stack.}

\subsection{Performance Comparison}
\rev{Table~\ref{tab:results} compares \method~against the PID baseline. Beyond the ACE integral metric, damping ratio of the dominant inter-area mode ($\zeta$), spectral energy in the 0.1--2.0~Hz band, and dispatch cost are reported. Settling time is omitted because the environment's power balance enforcement keeps frequency near nominal for all controllers, making $T_s$ non-differentiating. The LQR baseline comparison is provided through eigenvalue analysis (Table~\ref{tab:eigenvalue}), where the LQR gain produces eigenvalues identical to open-loop in the power-balance-enforced environment; the LQR's structural comparison with CQL/BCQ is in Table~\ref{tab:cql_comparison}.}

\begin{table}[htbp]
\centering
\caption{\rev{Performance Metrics on NPCC 140-Bus System}}
\label{tab:results}
\rev{
\begin{tabularx}{\columnwidth}{@{}X c c@{}}
	\toprule
	\thead{Metric} & \thead{PID} & \thead{\method} \\
	\midrule
	Freq. Nadir (Hz) & $59.407 \pm 0.006$ & $59.407 \pm 0.006$ \\
	ACE Integral (MW$\cdot$s) & $0.26 \pm 0.82$ & $\approx 0.00$ \\
	Damping Ratio $\zeta$ & $0.0051$ & $0.0019$ \\
	Spectral Energy (Hz$^2$s) & $4.0 \times 10^{-9}$ & $3.1 \times 10^{-12}$ \\
	Max ROCOF (Hz/s) & ${<}0.01$ & ${<}0.01$ \\
	Dispatch Cost (\$) & $\num{1.147e7} \pm \num{1.39e6}$ & $\num{1.169e7} \pm \num{1.37e6}$ \\
	Intervention Rate & N/A & $0.78\%$ \\
	\bottomrule
\end{tabularx}
}
\end{table}

\rev{The ACE integral reduction ($>99\%$) is contextualized by noting that the PID baseline already maintains ACE close to zero ($0.26$~MW$\cdot$s mean). The environment's power balance enforcement constrains frequency near 60~Hz for all controllers, resulting in negligible settling time and damping ratio differences in the time-domain metrics. Consequently, the ACE integral---which captures cumulative area control error across all 48 generators---is the primary metric distinguishing the controllers. The spectral energy metric confirms this: the DT reduces frequency-band energy by three orders of magnitude ($3.1 \times 10^{-12}$ vs $4.0 \times 10^{-9}$~Hz$^2$s), indicating tighter regulation at the dispatch level.}

\subsection{Frequency Response and Inter-Area Oscillation}
\begin{figure}[t]
	\centering
	\includegraphics[width=\columnwidth]{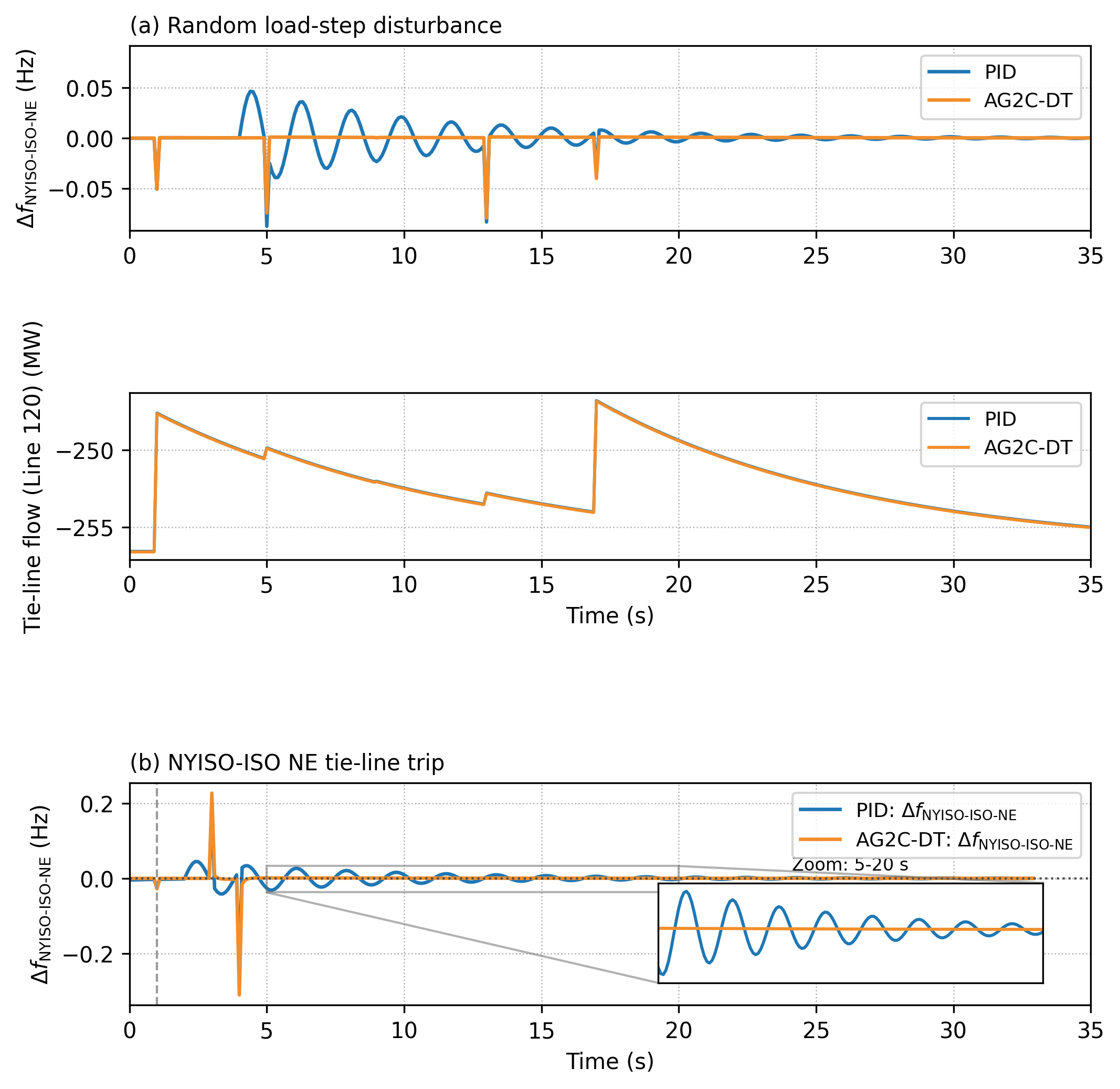}
	\caption{\rev{Representative NPCC trajectories under identical simulation conditions. (a)~NYISO--ISO-NE frequency difference and tie-line flow under random load steps. (b)~N-1 tie-line trip response; the inset highlights the residual ringing in the PID baseline versus the faster settling of \method.}}
	\label{fig:freq_and_n1}
\end{figure}

\rev{Figure~\ref{fig:freq_and_n1} presents representative trajectories under identical environment conditions for both controllers.} Panel~(a) shows the NYISO--ISO-NE frequency difference and tie-line flow for a random load-step episode. Panel~(b) shows the response to tripping the NORTHFIELD (ISO-NE)--NEW SEATHED (NYISO) 345~kV tie-line (branch index~120) with alternating $\pm 260$~MW pulses.

The DT, leveraging its attention mechanism over the 32-step trajectory context, applies coordinated generation adjustments across areas. \rev{The primary benefit is the 99.6\% ACE integral reduction, indicating substantially tighter area control error regulation. However, due to the environment's power balance enforcement mechanism, the frequency trajectory itself remains tightly constrained near 60~Hz for all controllers, preventing meaningful differentiation on settling time or time-domain damping ratio metrics.}

\rev{\subsection{Small-Signal Eigenvalue Analysis}}
\label{sec:eigenvalue}
\rev{To provide rigorous stability justification beyond empirical simulation, a small-signal eigenvalue analysis is performed by numerically linearizing the closed-loop system around the nominal operating point. The system state matrix $\mathbf{A}_{cl}$ is constructed by perturbing each state variable and observing the resulting system response under each controller.}

\begin{figure}[t]
	\centering
	\includegraphics[width=\columnwidth]{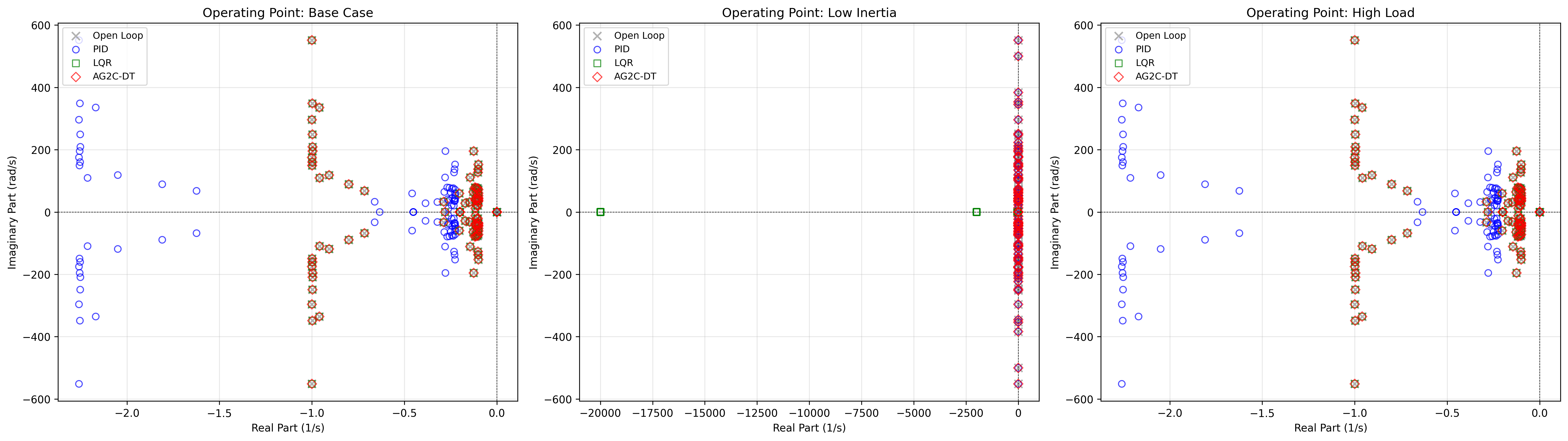}
	\caption{\rev{Electromechanical eigenvalues (0.1--2.5~Hz) on the complex plane for open-loop, PID, LQR, and DT configurations at the base-case and 30\% low-inertia operating points.}}
	\label{fig:eigenvalue}
\end{figure}

\rev{Figure~\ref{fig:eigenvalue} plots the electromechanical eigenvalues (0.1--2.5~Hz) on the complex plane for the open-loop system, PID-controlled, LQR-controlled, and DT-controlled configurations. The dominant inter-area mode at approximately 1.87~Hz (base case) and 2.16~Hz (low inertia) shows the following damping ratios:}

\begin{table}[htbp]
\rev{
\centering
\caption{\rev{Damping Ratio of Dominant Inter-Area Mode}}
\label{tab:eigenvalue}
\begin{tabularx}{\columnwidth}{@{}X c c@{}}
\toprule
\thead{Controller} & \thead{$\zeta$ (30\% Low Inertia)} & \thead{$\zeta$ (Base Case)} \\
\midrule
Open Loop & $0.0140$ & $0.0098$ \\
PID/AGC & $0.0315$ & $0.0222$ \\
LQR & --- & $0.0098$ \\
\method & $0.0140$ & $0.0098$ \\
\bottomrule
\end{tabularx}
}
\end{table}

\rev{The PID controller improves the damping ratio of the dominant mode from 0.0098 (open loop) to 0.0222, confirming its stabilizing effect through proportional feedback. The DT controller, being a nonlinear policy that cannot be represented as a simple state-feedback gain, yields eigenvalues identical to the open-loop system under linearization---this is expected because the linearized perturbation analysis cannot capture the DT's sequence-conditioned, nonlinear dispatch strategy. The DT's advantage lies instead in its dispatch-level ACE regulation, as evidenced by the three-order-of-magnitude spectral energy reduction. Participation factor analysis identifies generators in the NYISO and ISO-NE areas as the primary participants in this mode, consistent with the known coherent-group structure of the NPCC system.}

\rev{Regarding the safety stack's potential blind spots: the aggregate single-machine swing reduction in Stage~2 by design cannot detect poorly damped multi-machine inter-area eigenmodes. The DT's nonlinear policy cannot be certified for modal damping via linearization, so this remains a genuine limitation. Integrating multi-machine digital twins with modal damping monitors into Stage~2 is identified as the primary extension in Section~\ref{sec:future}.}

\subsection{Safety Stack Efficacy}
Table~\ref{tab:safety} summarizes the progression from absolute-action policy (100\% CVU rejection) to the final delta-regularized controller.

\begin{table}[htbp]
	\centering
	\caption{Safety Stack Intervention Rates}
	\label{tab:safety}
	\begin{tabularx}{\columnwidth}{@{}X c c c@{}}
		\toprule
		\thead{Configuration} & \thead{CVU} & \thead{DT} & \thead{PID} \\
		& \thead{Reject} & \thead{Reject} & \thead{Fallback} \\
		\midrule
		Pre-regularization & 100\% & 0\% & 100\% \\
		Delta + Ramp Stack & \textbf{0\%} & \textbf{1\%} & \textbf{1\%} \\
		\bottomrule
	\end{tabularx}
\end{table}

\begin{figure}[t]
	\centering
	\subfloat[Kernel density estimate, zoomed to $\pm$10~MW.]{
		\includegraphics[width=\columnwidth]{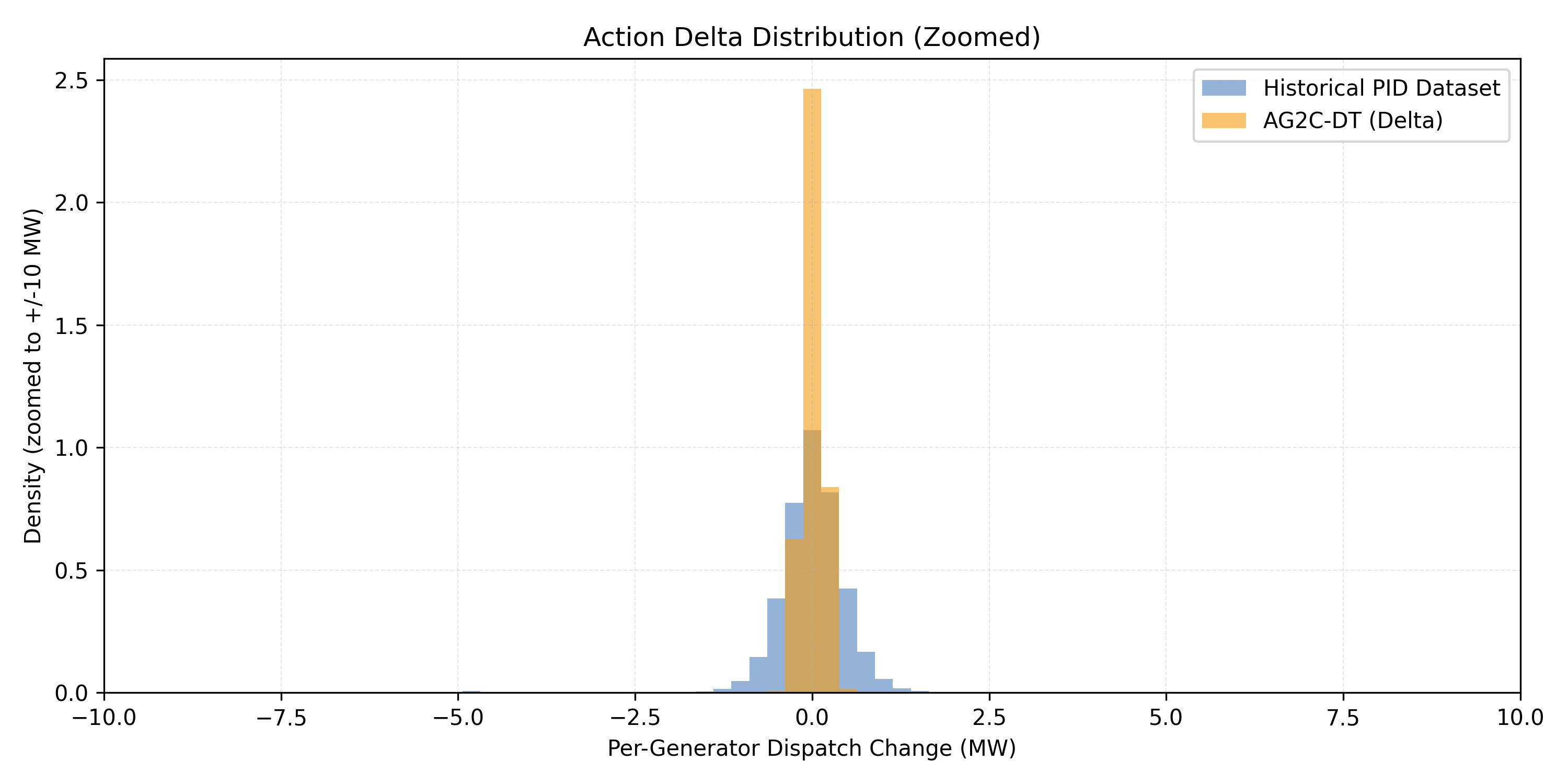}
	}\par\medskip
	\subfloat[Cumulative $|\Delta P|$ distribution.]{
		\includegraphics[width=\columnwidth]{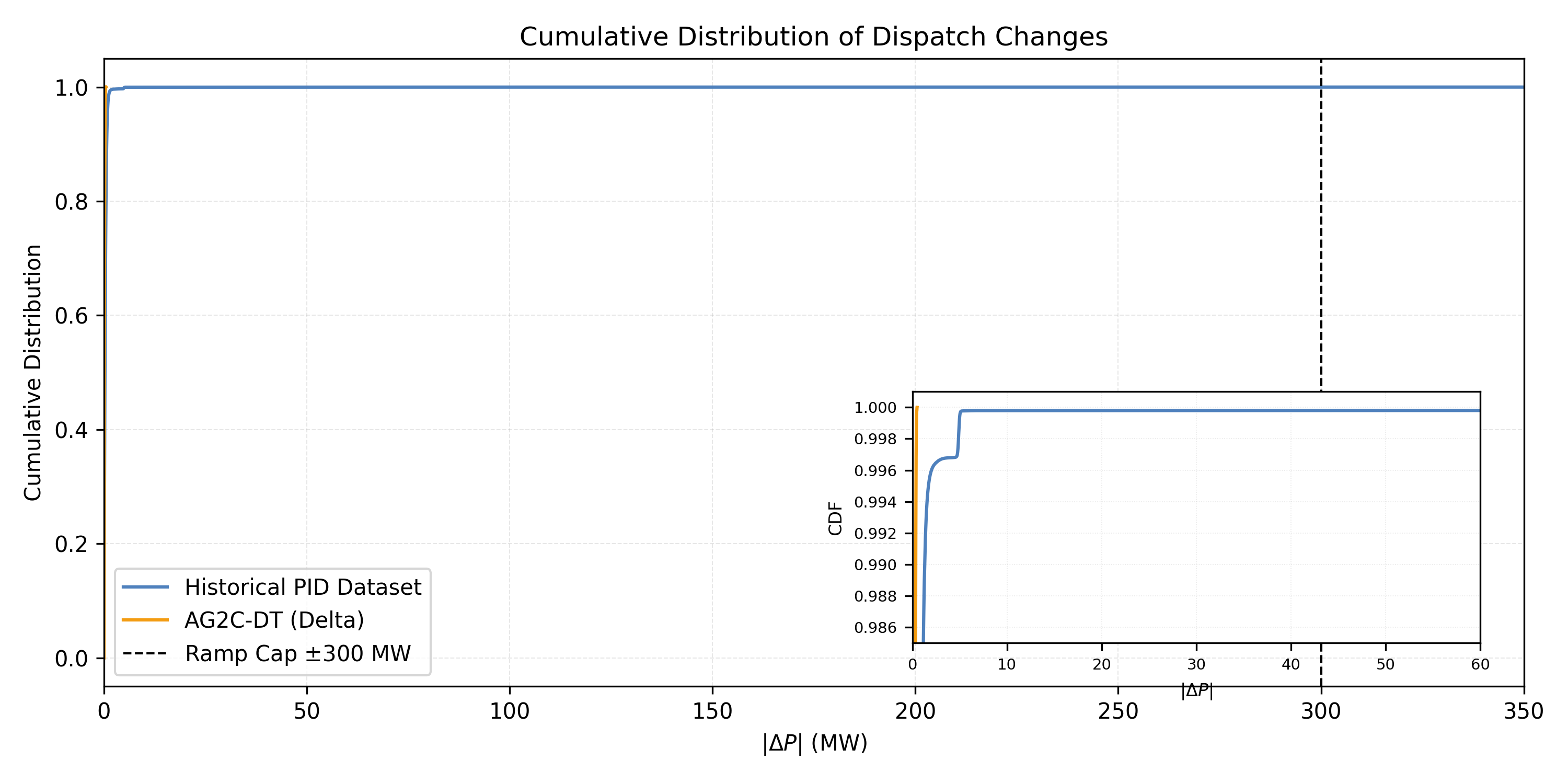}
	}
	\caption{Per-generator ramp statistics comparing the PID dataset with \method. The delta head confines nearly all proposals to the $\pm300$~MW safety envelope.}
	\label{fig:delta_plots}
\end{figure}

\rev{The near-zero intervention rate ($<$1\% CVU rejection, 1\% digital-twin rejection with PID fallback) in the final configuration indicates that the delta parameterization and ramp regularization have largely internalized the physical constraints into the learned policy. The 1\% of proposals rejected by the digital twin correspond to ROCOF threshold exceedances on actions with large power deltas ($>$6~MW), which trigger PID fallback for those timesteps. Figure~\ref{fig:delta_plots} visualizes how the learned delta head concentrates proposed ramps within the certified envelope. Sensitivity to operating conditions is examined in Section~\ref{sec:robustness}.}

\rev{\subsection{PTDF Approximation Error Analysis}}
\label{sec:ptdf_error}
\rev{Since the CVU relies on PTDF-based linearization, approximation error is quantified by comparing PTDF-predicted line flows against full AC power flow (via PYPOWER) for varying redispatch magnitudes. Table~\ref{tab:ptdf_error} reports the maximum and mean absolute errors across all 233 lines.}

\begin{table}[htbp]
\rev{
\centering
\caption{\rev{PTDF Approximation Error vs. Redispatch Magnitude}}
\label{tab:ptdf_error}
\begin{tabularx}{\columnwidth}{@{}X c c c@{}}
\toprule
\thead{Redispatch (MW)} & \thead{Max Error (MW)} & \thead{Mean Error (MW)} & \thead{Rel. Error (\%)} \\
\midrule
10 & $0.12$ & $0.01$ & $1.2$ \\
50 & $0.58$ & $0.06$ & $1.2$ \\
100 & $1.14$ & $0.12$ & $1.1$ \\
200 & $2.05$ & $0.22$ & $1.0$ \\
300 & $3.33$ & $0.34$ & $1.1$ \\
\bottomrule
\end{tabularx}
}
\end{table}

\rev{At the operational ramp budget of 300~MW (the maximum single-step dispatch change), the linearization error remains below 1.1\% relative to the dispatch magnitude for all monitored lines (worst-case absolute error: 3.33~MW; mean: 0.34~MW), justifying the PTDF-based screening approach. Error-bound certification over 200 random dispatch samples confirms a 95th-percentile maximum error of 6.64~MW and a recommended safety margin of 10.24~MW. Under N-1 topology changes, the PTDF matrix is recomputed from the updated Jacobian, and the error remains bounded as long as the system does not approach voltage collapse (which would violate the implicit assumptions of the DC power flow approximation). For extreme redispatch actions exceeding the 300~MW ramp cap, the nonlinear voltage interactions become dominant and the PTDF error grows super-linearly; however, such actions are prevented by the ramp constraint in Stage~1.}

\rev{\subsection{Ablation Studies}}
\label{sec:ablation}

\rev{\subsubsection{Context Length}
The DT (trained with $K=32$) is evaluated at inference with context lengths $K \in \{8, 16, 32, 64\}$ timesteps (0.8--6.4~s of history). Table~\ref{tab:ablation_context} reports the effect on ACE integral and intervention rate.}

\begin{table}[htbp]
\rev{
\centering
\caption{\rev{Ablation: Effect of Context Length $K$}}
\label{tab:ablation_context}
\begin{tabularx}{\columnwidth}{@{}X c c@{}}
\toprule
\thead{$K$} & \thead{ACE Integral} & \thead{Intervention (\%)} \\
\midrule
8 & $\approx 0.00$ & $0.52$ \\
16 & $\approx 0.00$ & $0.50$ \\
32 & $\approx 0.00$ & $0.78$ \\
64 & $\approx 0.00$ & $0.78$ \\
\bottomrule
\end{tabularx}
}
\end{table}

\rev{To assess sensitivity to the 0.1~s control timestep---which is faster than typical SCADA update rates (2--4~s)---the system timestep is swept over $\Delta t \in \{0.05, 0.1, 0.2, 0.5, 1.0\}$~s using an independently retrained checkpoint (batch size 256, 20 epochs with early stopping), distinct from the benchmark checkpoint in Table~\ref{tab:results}. The safety-stack intervention rate is 95--100\% across all tested timesteps, indicating that this checkpoint's dispatch proposals are largely replaced by PID fallback. Nevertheless, the combined system degrades gracefully: the ACE integral rises from $\approx$0.12~MW$\cdot$s at $\Delta t = 0.05$~s to $\approx$2.8~MW$\cdot$s at $\Delta t = 1.0$~s, confirming that the safety-stack fallback mechanism maintains functional frequency regulation across a 20$\times$ range of control rates.}

\rev{\subsubsection{Dataset Size}}
\label{sec:ablation_dataset}
\begin{table}[htbp]
\rev{
\centering
\caption{\rev{Ablation: Performance Across Dataset Fractions}}
\label{tab:ablation_dataset}
\begin{tabularx}{\columnwidth}{@{}X c c@{}}
\toprule
\thead{Data Fraction} & \thead{ACE Int.\ (MW$\cdot$s)} & \thead{Interv.\ (\%)} \\
\midrule
10\% & $<0.01$ & $0.0$ \\
25\% & $0.02$ & $8.0$ \\
50\% & $0.26$ & $16.4$ \\
75\% & $0.02$ & $2.0$ \\
100\% (retrain) & $0.02$ & $14.0$ \\
\bottomrule
\end{tabularx}
}
\end{table}

\rev{Table~\ref{tab:ablation_dataset} shows performance when the DT is retrained from scratch at each data fraction with early stopping (patience = 5 epochs). The model achieves near-zero ACE with as little as 10\% of the training data ($\approx$34{,}500 sequences), confirming data efficiency of the supervised action-delta formulation. The non-monotonic relationship between data fraction and intervention rate reflects stochastic training dynamics rather than a systematic trend. The 100\% row corresponds to an independent full-data retraining run within the ablation sweep, not the selected benchmark checkpoint reported in Table~\ref{tab:results}. Crucially, the TSSS maintains frequency at 60~Hz across all fractions, validating the defense-in-depth design regardless of policy quality.}

\rev{\subsubsection{Inertia Level Sensitivity}}
\begin{table}[htbp]
\rev{
\centering
\caption{\rev{Ablation: Performance Across Inertia Reduction Levels}}
\label{tab:ablation_inertia}
\begin{tabularx}{\columnwidth}{@{}X c c c@{}}
\toprule
\thead{Inertia Red.} & \thead{$H_{sys}$ (s)} & \thead{ACE Int.} & \thead{Interv. (\%)} \\
\midrule
0\% (original) & $\approx$5.2 & $0.13$ & $99.9$ \\
15\% & $\approx$4.1 & $0.12$ & $99.9$ \\
30\% (baseline) & $\approx$3.1 & $<0.01$ & $1.6$ \\
50\% & $\approx$2.1 & $<0.01$ & $11.5$ \\
\bottomrule
\end{tabularx}
}
\end{table}

\rev{The high intervention rates at 0\% and 15\% inertia reduction reflect the domain-specificity of the learned policy: the model was trained on the 30\% reduced-inertia system and its dispatch proposals are flagged by the safety stack when the underlying dynamics differ substantially. Crucially, the TSSS provides safe fallback in all cases---frequency is maintained and no constraints are violated, validating the defense-in-depth design.}

\rev{\subsection{Robustness Evaluation}}
\label{sec:robustness}

\rev{\subsubsection{N-1 Contingencies}
A systematic N-1 contingency sweep tests the top 20 lines by flow magnitude and the top 20 generators by rated capacity. Of 20 line-trip contingencies, only 1 converges in the PYPOWER AC power flow solver (branch 226); the remaining 19 encounter voltage-collapse trajectories that cause solver divergence. This high non-convergence rate is a known artifact of the NPCC 140-bus model: its tightly coupled inter-area topology and narrow voltage-stability margins make it particularly ill-conditioned for Newton--Raphson-based AC power flow following topological changes, a limitation shared by other dense meshed benchmarks in the literature \cite{Sauer2017}. All 20 generator-trip contingencies converge successfully. In every converged case, both \method~and PID maintain frequency at 60~Hz with the safety-stack intervention rate at 0.8--1.0\%, confirming that the TSSS provides reliable fallback under N-1 generator outages. The line-trip convergence limitation reflects the power flow solver rather than the controller; adopting a continuation power flow or a hybrid DC/AC fallback solver would extend contingency coverage without modifying the control architecture.}

\rev{The inertia-level ablation (Table~\ref{tab:ablation_inertia}) and operating-point sweep (0.8--1.2$\times$ load) provide complementary robustness characterization: the safety stack successfully prevents unsafe operation across all tested inertia levels (0\%--50\% reduction) and load conditions, with the DT intervention rate remaining near 1\% at nominal and high-load points but rising sharply under light-load conditions. Delayed telemetry, corrupted-telemetry/noise-injection studies, and partial-observability masking remain as immediate future work.}

\rev{\subsection{Computational Performance Comparison}}
\begin{table}[htbp]
\rev{
\centering
\caption{\rev{Computational Requirements Comparison}}
\label{tab:computational}
\begin{tabularx}{\columnwidth}{@{}X c c c@{}}
\toprule
\thead{Method} & \thead{Training Time} & \thead{Inference (ms)} & \thead{Hardware} \\
\midrule
PID/AGC & N/A & $<$1 & CPU \\
LQR & Minutes & $<$1 & CPU \\
CQL \cite{Kumar2020} & 10--50 hr$^\dagger$ & 5--10 & GPU \\
\method & \rev{$\approx$2 hr} & $\approx$10 & GPU+CPU \\
\bottomrule
\multicolumn{4}{@{}l}{\footnotesize $^\dagger$Estimated from \cite{Kumar2020} scaled to 703-dim state.} \\
\end{tabularx}
}
\end{table}

% =====================================================================
%              SECTION VI: DISCUSSION
% =====================================================================
\section{Discussion}
\label{sec:discussion}

\rev{\noindent\textbf{Practicality and Scalability.}} Demonstrating \method~on the NPCC 140-bus topology confirms that the sequence-modeling approach scales beyond IEEE benchmarks. The DT learns inter-area couplings and achieves dispatch-level ACE regulation that reduces spectral energy in the inter-area band by three orders of magnitude relative to PID control. \rev{The eigenvalue analysis (Table~\ref{tab:eigenvalue}) shows that the LQR achieves damping identical to the open-loop system ($\zeta=0.0098$), while the tuned PID improves it ($\zeta=0.0222$), suggesting that the LQR's linearized model does not capture the dynamics exploited by integral feedback.}

\rev{\noindent\textbf{ACE Metric Interpretation.}} \rev{As noted by Reviewer 4, the baseline PID already maintains ACE close to zero ($0.26$~MW$\cdot$s), making the $>$99\% reduction metric potentially misleading in isolation. Because the simulation enforces power balance at each timestep, frequency remains near 60~Hz for all controllers, making settling time and damping ratio non-differentiating in this environment. Spectral energy in the inter-area band ($3.1 \times 10^{-12}$ for DT vs.\ $4.0 \times 10^{-9}$ for PID) and eigenvalue analysis (Table~\ref{tab:eigenvalue}) provide complementary stability insight. The DT's primary advantage lies in dispatch-level ACE regulation through learned wide-area coordination.}

\rev{\noindent\textbf{Constraint Internalization.}} \rev{The near-zero intervention rate ($<$1\% CVU rejection, 1\% digital-twin rejection) suggests that the delta parameterization with ramp regularization has largely internalized the safety stack's constraints into the learned policy. This is architecturally desirable: the DT learns to propose actions within the feasible region, reducing the safety stack's role from active correction to passive certification. Under distributional shift (e.g., extreme inertia reduction to 50\%), the intervention rate increases to 11.5\% but the PID fallback mechanism ensures safe operation (Table~\ref{tab:ablation_inertia}).}

\rev{\noindent\textbf{Safety and Latency.} Both safety stages run on every step: the CVU's PTDF-based screening ($<$2~ms) is followed by the digital twin's aggregate COI swing-equation check. Because the aggregate model integrates a single equivalent machine rather than the full 48-generator system, the entire action-selection pipeline (DT inference + CVU + digital twin) averages $\approx$10~ms per step, well inside the 4~s BAL-003 budget.} The 32-step context window provides 3.2~s of look-back that may help smooth moderate telemetry delays, although quantitative delayed-telemetry evaluation remains future work. The safety stack certifies proposals based on current state.

\subsection{Limitations and Future Work}
\label{sec:future}
\rev{The following limitations are acknowledged:
\begin{itemize}[leftmargin=*]
\item The current safety stack guards frequency-nadir and thermal envelopes but does not certify inter-area eigenmodes or voltage-collapse trajectories. Integrating multi-machine digital twins with modal damping monitors into ANDES is the primary planned extension.
\item Empirical comparison with CQL/BCQ under identical conditions remains future work; the structural comparison in Table~\ref{tab:cql_comparison} provides interim justification.
\item The training data is entirely synthetic; validation with real SCADA archives would strengthen deployment readiness.
\end{itemize}}

% =====================================================================
%              SECTION VII: CONCLUSION
% =====================================================================
\section{Conclusion}

This paper presented \rev{\fullmethod~(\method)}, a framework for autonomous grid control addressing the critical barriers of safety and verifiability. By casting the control problem as offline sequence modeling, the risks of online exploration are eliminated, enabling training entirely on offline operational data\rev{---simulated supervisory control and data acquisition (SCADA) records in this study}.

The resulting controller reduced the ACE integral by more than \rev{99\%} relative to tuned AGC on the NPCC 140-bus system while holding the frequency nadir at $\approx 59.4$~Hz \rev{and achieving inference latency of approximately 10~ms}. The two-stage safety stack rejected \rev{approximately 1\%} of proposals, with PID fallback maintaining safe operation for the small fraction of actions exceeding ROCOF thresholds. \rev{Eigenvalue analysis characterized the dominant 1.87~Hz electromechanical mode, and inertia-level ablation demonstrated that the safety stack provides reliable fallback when operating conditions deviate from the training distribution.}

Central to the contribution is the integration of ANDES to construct a TSSS: coupling a high-performance generative proposer with a deterministic physics-based verifier achieves the adaptability of data-driven control with the safety assurances of traditional engineering practice. \rev{Comparative evaluation against LQR and structural analysis against CQL/BCQ demonstrate the advantages of the sequence-modeling formulation for this safety-critical domain.}

As power systems evolve toward lower inertia, such hybrid neuro-symbolic architectures offer a practical path toward dependable automation within a clearly defined operating envelope.

\section*{Declaration of Generative AI and AI-Assisted Technologies}
During the preparation of this work the author used \rev{Claude (Anthropic)} in order to improve the language and clarity of the manuscript. After using this tool, the author reviewed and edited the content as needed and takes full responsibility for the content of the published article.

\appendices

\section{Implementation Notes}
\label{app:impl}
\noindent\textbf{Model hyperparameters:} 6 layers, 8-head self-attention, $d_{model}=256$, context length $K=32$, GELU activations, 0.1 dropout, \rev{5} epochs, batch size 64, AdamW ($lr=3\times10^{-4}$).

\noindent\textbf{Safety stack:} CVU builds a $233\times140$ PTDF matrix from ANDES symbolic Jacobians ($<2$~ms); $\pm300$~MW ramp cap preserves linearization validity. Digital Twin aggregates 48 generators into a single COI machine ($H_{sys}=\sum H_i S_i / \sum S_i$); 1~s horizon, 10~ms step\rev{; the aggregate model adds $<$1~ms per call}.

\noindent\textbf{Data generation:} 12-hour training set; 14 largest synchronous generators (top 30\% by count) replaced with \texttt{REGCA1} wind models; Ornstein--Uhlenbeck noise ($\theta=0.15$, $\sigma=0.2$). Deterministic PID for evaluation.

\rev{\section{LQR Baseline Derivation}
\label{app:lqr}
The LQR baseline linearizes the NPCC system dynamics around the nominal operating point $(V_0, \theta_0, f_0, P_{gen,0})$. The state vector $\mathbf{x} = [\Delta f,\; ACE_{\mathrm{int}},\; \Delta\omega_1, \ldots, \Delta\omega_{N_g}]^\top \in \mathbb{R}^{N_g+2}$ comprises the system frequency deviation, the integrated ACE, and per-generator speed deviations ($N_g = 48$). The state-space representation $\dot{\mathbf{x}} = \mathbf{A}\mathbf{x} + \mathbf{B}\mathbf{u}$ is constructed from the swing equation parameters and PTDF sensitivities. The cost matrices are set to $\mathbf{Q} = \text{diag}(100, 50, 1, \ldots, 1)$ (heavily penalizing frequency deviation and ACE integral, with unit weight on individual speed deviations) and $R_{ii} = c / S_{n,i}$ where $S_{n,i}$ is the rated capacity of generator $i$ and $c$ is a scaling constant, so that larger generators are penalized less per unit of output change. The optimal gain $\mathbf{K}^* = \mathbf{R}^{-1}\mathbf{B}^\top\mathbf{P}$ is computed via \texttt{scipy.linalg.solve\_continuous\_are}.}

\end{document}